\documentclass[11pt,twocolumn]{article}
\usepackage{graphicx}
\usepackage{algorithm}
\usepackage{algorithmic}
\usepackage[T1]{fontenc}
\usepackage{ae,aecompl}
\usepackage[super]{natbib}

\begin{document}
\title{\Large{{\bf Improved RNA pseudoknots prediction and classification using a new
topological invariant}}}
\vspace{1.5cm}
\renewcommand\thefootnote{\relax}
\author{ 
{\sc Graziano Vernizzi}$^{1}$ ,
{\sc Henri Orland}$^{2,3}$ and
{\sc A. Zee}$^{4,5}$ \\
$^1$ Department of Physics and Astronomy, Siena College, New York, USA \\
$^2$Institut de Physique Th\'eorique, CEA Saclay, 91191
Gif-sur-Yvette Cedex, France\\
$^3$Beijing Computational Science Research Center, Haidian District Beijing, 100084, China \\
$^4$Department of Physics, University of California, Santa Barbara,
CA 93106, USA\\
$^5$Kavli Institute for Theoretical Physics, University of
California, Santa Barbara, CA 93106, USA\\
}
 
\twocolumn[
\begin{@twocolumnfalse}
\maketitle

\begin{abstract}
We propose a new topological characterization of RNA
secondary structures with pseudoknots based on two topological invariants. Starting  from the classic arc-representation of RNA secondary structures, we consider a model that couples both I) the topological genus of the graph and II) the number of crossing arcs of the corresponding primitive graph. We add a term proportional to these topological invariants to the standard free energy of the RNA molecule, thus obtaining  a novel free energy parametrization which takes into account the abundance of topologies of RNA pseudoknots observed in RNA databases. 

 \vfill
    \end{abstract}
  \end{@twocolumnfalse}
]
\footnotetext[1]{{\it Email addresses:}
\texttt{gvernizzi@siena.edu} (Graziano Vernizzi),
\texttt{Henri.Orland@cea.fr} (Henri Orland),
\texttt{zee@kitp.ucsb.edu} (A. Zee).} \footnotetext[1]{$^*$All
authors contributed equally to this work.}

The prediction of possible foldings of RNA molecules is still a major open problem of molecular biology \cite{1,2}. It is of utmost importance, since the three-dimensional structure of any folded biopolymer mostly determines its biological function by providing the adequate geometry for biochemical reactions to occur. In the last thirty years, the role of RNA has been upgraded from being a relatively minor player in the central dogma of Watson and Crick to being one of the central players in molecular biology \cite{1}. It has been recognized that in addition to being a carrier of genetic information, some RNA may also have enzymatic roles, and may play a central part in the regulation of biological networks \cite{1}. In spite of considerable effort, the accurate prediction of the three-dimensional structure of RNA from its primary sequence has resisted so far the most advanced computational methods, in particular for long RNA sequences. In such cases, drastic approximations are necessary. A typical simplifying assumption is that the RNA secondary structure (i.e. the complete list of paired nucleotides) already provides sufficient information on the active sites of the RNA molecules, by allowing the identification of loops and other motifs such as pseudoknots, where the biochemistry takes place \cite{3}. The energetic landscape of an RNA molecule is mostly dominated  by Crick-Watson base pairings (A,U), (G,C), and the additional wobble pair (G,U). Non-canonical base pairs and tertiary interactions have been recognized to further stabilize the structure\cite{4} of RNA, nonetheless we will not include them in the present work. Several deterministic and stochastic methods have been proposed for the prediction of secondary structures of RNA molecules \cite{5,6,7,8}. Despite great progress, their overall success is limited, in particular for long RNA molecules. Part of the difficulty lies in the prediction of RNA pseudoknots, which has been identified as an $NP$-complete problem \cite{9}.

\begin{figure}[h]
  \centering
   \includegraphics[width=3.2in]{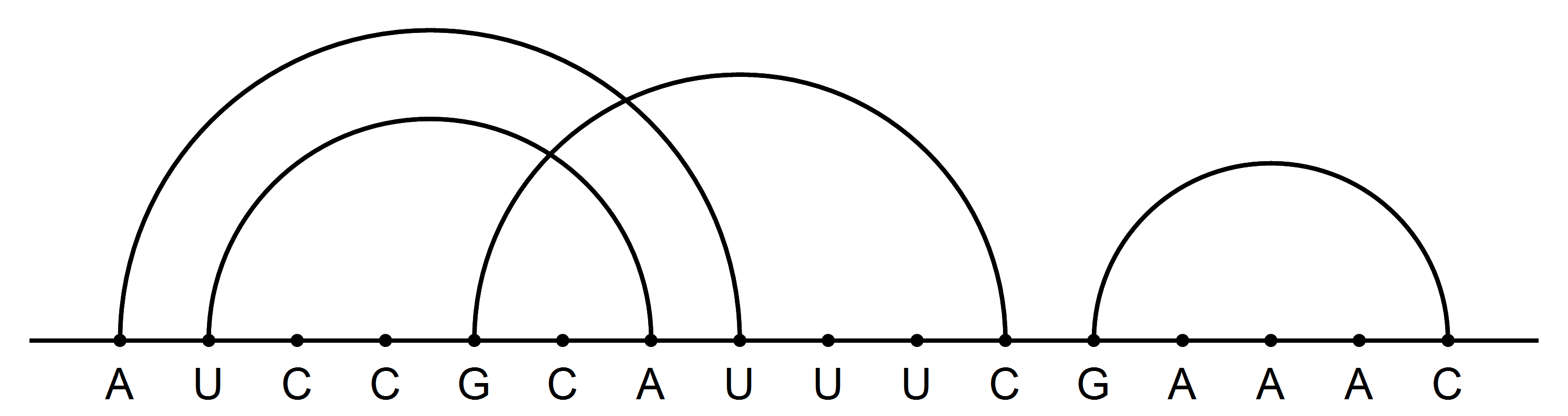}
     \put(-230,12){$5'$}
     \put(-10,12){$3'$}
  \caption{In the arc-diagram representation of an RNA, crossing arcs indicate the presence of a pseudoknot.}
  \label{f1}
\end{figure}

We now summarize some standard notations to represent all base-pairings in a RNA molecule. 
The backbone of an RNA molecule can be represented by an oriented straight line (from the $5'$ to the $3'$ end), on which the nucleotides appear in the order given by the RNA primary sequence. A pairing between two bases is depicted by an arc joining the two bases in the upper half-plane above the backbone line (see Fig.~\ref{f1}). A graph without crossing pairing lines is called a {\it planar graph}. If a graph contains lines that cross, then it is said to contain a {\it pseudoknot}.
If one assigns a suitable pairing energy (called {\it stacking energy}) to adjacent base pairs, then it is possible to compute the partition function of all planar graphs exactly, by using standard recursion equations \cite{11,12}. 
However, if one allows for the occurrence of pseudoknots  then those recursive algorithms face an overwhelming increase in polynomial complexity. Several alternative algorithms have been proposed to predict pseudoknotted structures \cite{13,14,15,16,17}. 

\begin{figure}[h]
  \centering
   \includegraphics[width=3.2in]{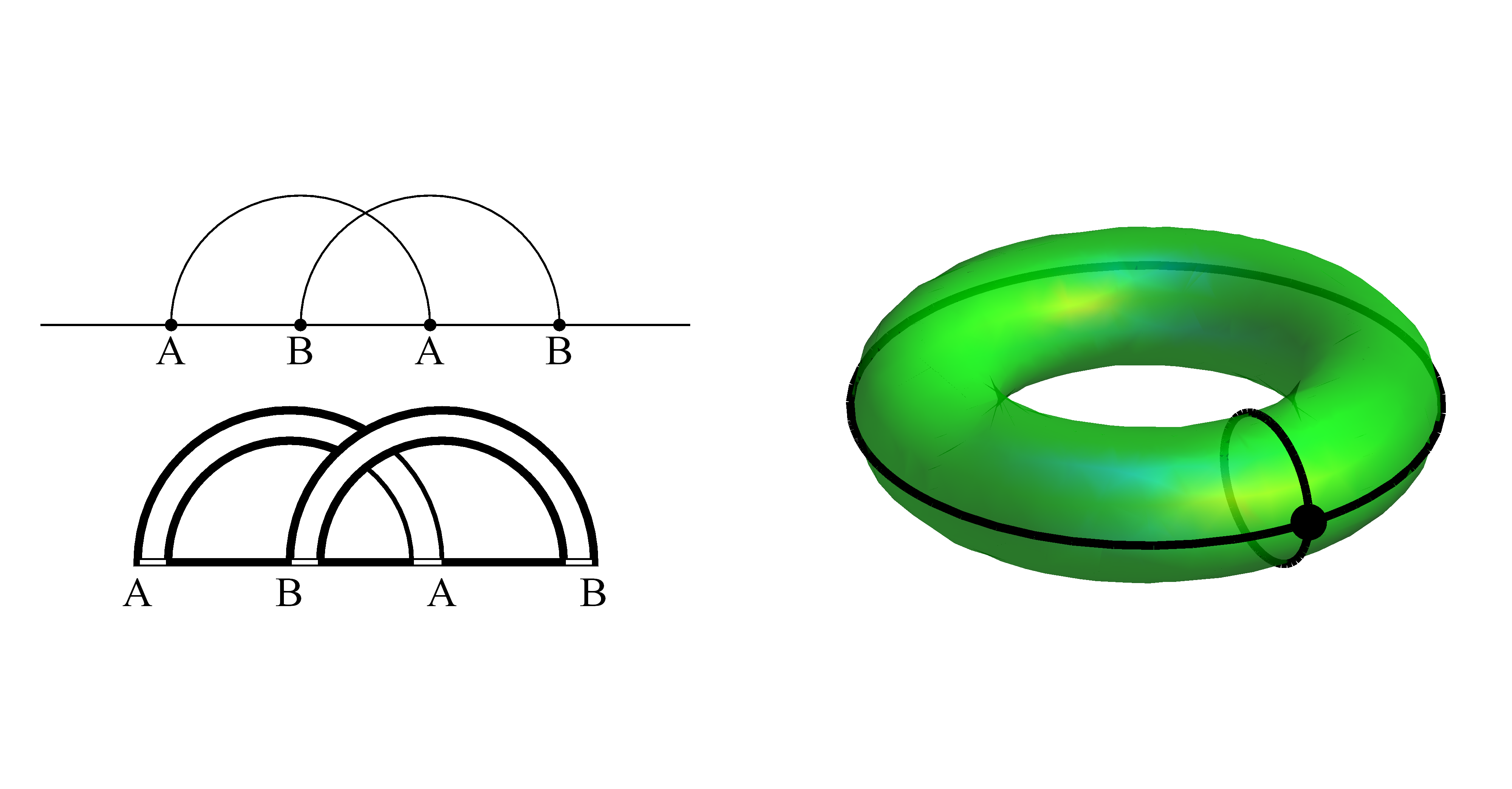}
  \caption{The arc-diagam representation of an ``ABAB'' H-pseudoknot, its double-line representation, and its embedding on a torus on which the arc-diagram can be drawn without crossings. This corresponds to a topological genus $g=1$. }
  \label{f2}
\end{figure}

We have proposed a topological classification of pseudoknots in terms of their genus \cite{16}, followed by two algorithms for the prediction of such pseudoknots \cite{17,18}. The genus of an RNA graph can be defined in the following way \cite{19}: join the $5'$-end with  the $3'$-end by bending the backbone line in the lower-half plane to make a circle, so that all pairing lines exist on the outside of the circle. The actual size of such a circle is of course irrelevant, and it is therefore topologically equivalent to a puncture on a surface. The genus of the graph is the minimal number of handles one has to carve in a punctured sphere, so that the graph can be drawn  on it without any crossing. A planar graph by definition can be drawn on a sphere without any crossing arc, and so it is of genus $g=0$ (the sphere has no handles). A H-pseudoknot (i.e. the ``ABAB'' pseudoknot with two  helices A and B) can be drawn without crossing on a torus, which is a sphere with one handle and therefore with genus $g=1$ (see Fig. \ref{f2}).

A practical diagrammatic way to compute the genus of a graph is by using the so-called double-line representation, where base-pairs are drawn using oriented double lines (see Fig. \ref{f2}). In such a representation, oriented loops appear on the graph. The genus can be shown to be equal to $g= (p-l)/2$ where $p$ is the number of pairings of the graph (i.e. the number of arcs) and $l$ is the number of closed loops. The genus allows to organize pseudoknots and secondary structures of RNA systematically in equivalence classes, each class corresponding to a value of the genus $g$ \cite{161,16}. It is a topological invariant which depends only on the connectivity of the RNA base-pairs. Moreover, it has the property of being additive: if a structure comprises two consecutive pseudoknots with genus $g_1$ and $g_2$, the genus of the whole RNA sequence is $g=g_1+g_2$. 
However, it is known experimentally that pseudoknots are fairly rare in RNA molecules \cite{161}. Furthermore, they usually impose some mechanical constraint on the sugar-phosphate backbone of the molecule. We have thus proposed\cite{17,18} to add an energetic  penalty  proportional to the genus, to the standard folding energy (which includes stacking energies, loop penalties, etc.).
Within such a framework, the partition function of the system is
 \begin{equation}
 \label{eq1}
 {\cal Z}=\sum_{all \, graphs } e^{-\beta \left[ E(graph) + \mu_g g (graph) \right] }
\end{equation}
where $\beta= 1/k_BT$ is the inverse temperature, $k_B$ is the Boltzmann constant, $E$ is the free energy (which phenomenologically includes the configurational entropy at fixed genus) and $g$ is the topological genus. The parameter $\mu_g$ is a phenomenological parameter, used to penalize graphs with high genus. Planar graphs, i.e. graphs without pseudoknots are obtained by taking $\mu_g$ to infinity
\cite{19,20}.

We have developed two algorithms to sample the partition function in eq.~(\ref{eq1}) and predict the secondary structures of RNAs with pseudoknots\cite{17,18}. In \cite{17}, we first make a library of possible paired RNA segments from the sequence. We then enumerate all the possible assemblies of these fragments and compute the corresponding free energy. The minimal free energy state can be computed, but the method is limited to fairly small sizes ($L<150$ where $L$ is the number of nucleobases). In \cite{18}, we start from the same library of building blocks, but we assemble them using a Monte Carlo algorithm (multiple Markov chains). This last method allows to handle RNAs of sizes up to 1000 nucleobases.

\begin{figure}[h]
  \centering
   \includegraphics[width=3in]{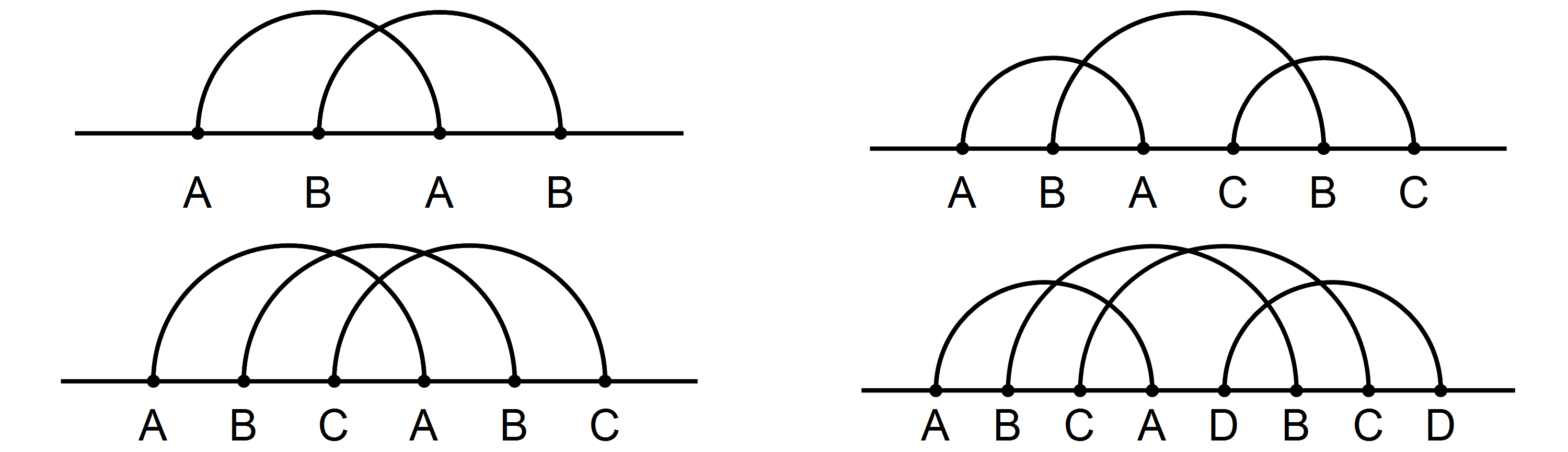}
  \caption{There are four type of primitive pseudoknots with genus $g=1$.}
  \label{f3}
\end{figure}

Although methods based on eq.~(\ref{eq1}) are promising, they do not predict correctly the abundance of various structures with identical genus. For example, following ref.\cite{161}, there are four primitive graphs of genus $g=1$. We define a {\it primitive graph} as a graph which is both irreducible (i.e. cannot be disconnected by cutting the backbone somewhere) and non-nested  (i.e. cannot be disconnected by cutting {\it twice} the backbone somewhere), and in which all equivalent parallel pairing arcs  (i.e. a  {\it sheaf} of parallel arcs) are collapsed into a single {\it renormalized} arc. Later in this paper, we give an alternative definition, but completely equivalent.  In Fig. \ref{f3}, we sketch all four primitive graphs with genus $g=1$ (which have been obtained first in ref.\cite{Pil05} by steepest descent methods). 

With obvious notations, the 4 pseudoknots can be labeled as ABAB, ABACBC, ABCABC, ABCADBCD. As it was shown in ref. \cite{161}, the abundance of ABAB, either in the databases PDB or in PseudoBase, is much larger than that of ABACBC. The ABCABC pseudoknot is quite rare while the ABCADBCD is absent from the databases. This variation in abundance of the various genus 1 primitive pseudoknots is hardly understandable if the energetic penalty is only dependent on the genus.  
To account for this variation within a given genus, it is thus necessary to supplement the free energy by an additional term which would further discriminate between the structures. If we look at the four graphs of fig. \ref{f3}, we see that they differ by the number of crossings (i.e. crossing arcs) of the effective pairing arches. The ABAB graph has 1 crossing, ABACBC has 2 crossings, ABCABC has 3 crossings and ABCADBCD has 5 crossings. It turns out that their abundance decreases as a function of the number of crossings. 

As the number of crossings of a primitive graph is an additive quantity, it is natural to include an energetic penalty proportional to this number. 
We therefore introduce the {\it renormalized crossing number} in the following way:
\begin{enumerate}
\item  Given a generic graph ${\cal D}$, let ${\cal D}={\cal D}_1+{\cal D}_2+\ldots$ be its decomposition in irreducible or nested parts ${\cal D}_i$.
\item For each graph ${\cal D}_i$ we consider its  primitive version ${\cal D}'_i$ (i.e. all stacked arcs are collapsed into a single renormalized arc).
\item The renormalized crossing number $N_c$ of ${\cal D}$ is defined as the sum of the crossing number of each  ${\cal D}'_i$.  
\end{enumerate}
Such a definition allows to generalize the free energy for a RNA graph:
\begin{equation}
E_1=E(graph) + \mu_g g(graph) + \mu_c N_c(graph) \, ,
\end{equation}
and
\begin{equation}
\label{eq2}
{\cal Z}=\sum_{all \, graphs} e^{-\beta E_1} \, ,
\end{equation}
where $N_c(graph)$ denotes the renormalized crossing number of a given $graph$, and $\mu_c$ controls the associated energetic penalty. As was shown in ref. \cite{17}, a typical value for the genus penalty is $\beta \mu_g = 1.5$. \\

\begin{figure}[h]
  \centering
   \includegraphics[width=3in]{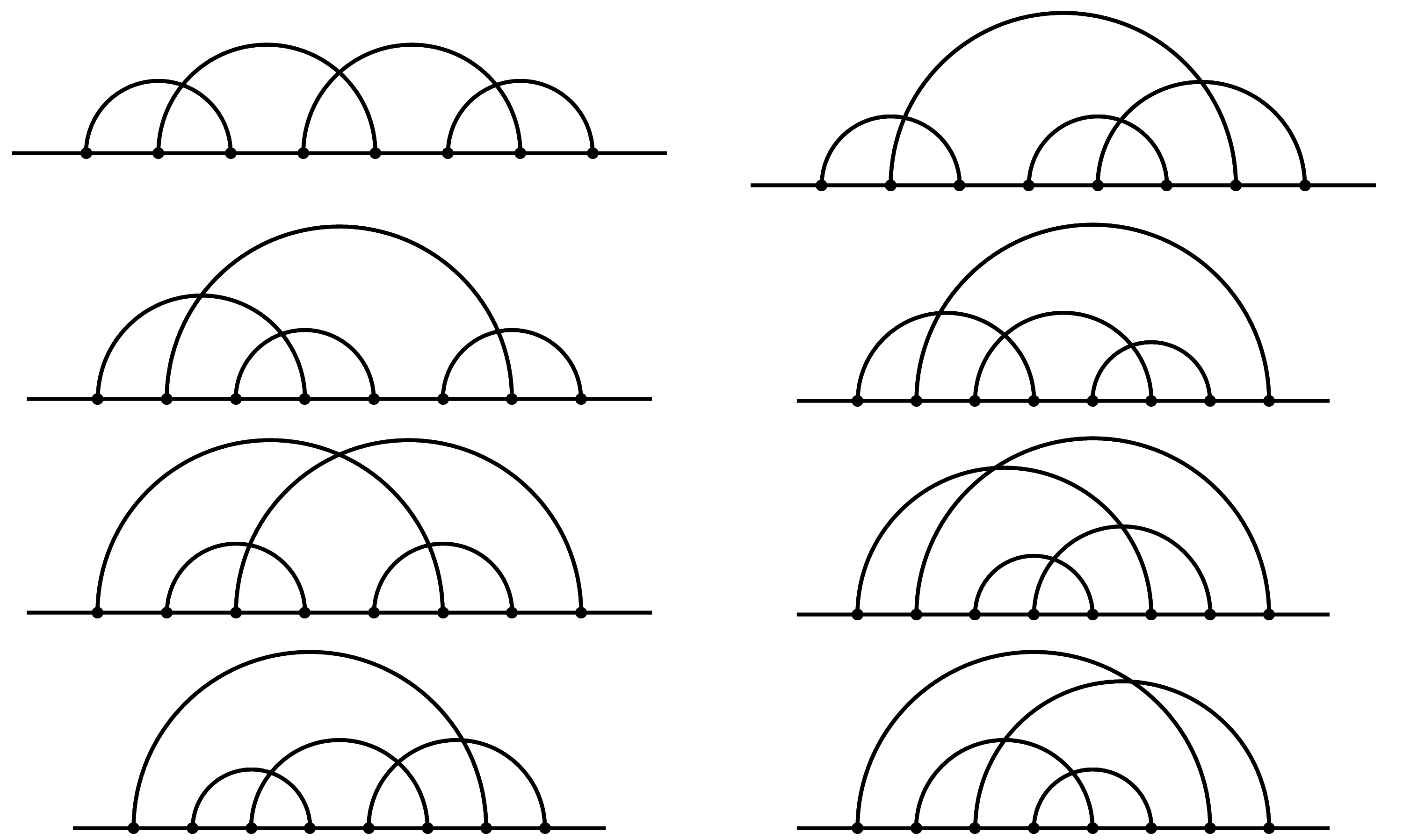}
  \caption{There are 8 primitive pseudoknots with genus $g=2$ and crossing number $N_c=3$.}
  \label{f6}
\end{figure}

The crossing penalty can be estimated by trying to fit the abundance of the various types of genus 1 pseudoknots. Currently, there are 398 pseudoknots in the {\it Pseudobase} database\cite{pseudobase}. In particular there are 355 ABAB graphs, 7 ABACBC graphs, 1  ABCABC graph, and no ABCADBCD graphs (for a total of  363 pseudoknots with genus 1). Such an ``exponential'' decay can be roughly described by using an approximate value of $\beta \mu_c \approx 2.5$. This provides a convenient way to account for the under-represented abundance of the ABCABC pseudoknot and the absence of the ABCADBCD pseudoknot, both of genus 1. It is worth emphasizing here that the relative abundance of different pseudoknot classes can be described by introducing a {\it single} linear term in the free energy, with a the corresponding phenomenological parameter $\mu_c$. Furthermore, in Pseudobase there are also 35  ABCDCADB graphs with genus 2. The latter represent a slightly biased sample since they {\it all} are of the  HDV-like ribozyme type (see the diagram on the second column, third  row of Figure \ref{f6}). A more systematic fit of the genus and number crossing penalties will be presented in a forthcoming study.

An important remark is in order at this point: the genus and the crossing number do not uniquely specify an RNA graph. Indeed, it is easy to see that except for $g=1$ there may exist several graphs with same genus and crossing number. In Figure \ref{f6} we display 8 graphs with genus $g=2$ and crossing number $N_c=3$.
Note that the introduction of a crossing penalty $\mu_c$ requires to recompute the value of the genus penalty $\mu_g$.
A more precise determination of both penalties will be performed in a forthcoming study, by optimizing them in order to improve the success rate of the prediction algorithms.

\section{Implementation and Algorithms}
In this section we describe some algorithms to a) extract the primitive graph from any RNA diagram, b) to compute its genus and c) its renormalized crossing number. For practical software implementations it is convenient to represent the pairing of a generic RNA diagram by using a formalism based on permutations. Given an RNA sequence with $L$ bases, each base can be identified by an integer number $i=1,\ldots,L$, from the $5'$ end to the $3'$ end. A specific pairing $(i,j)$ is denoted by a permutation $\pi$ with  $\pi(i)=j$. Obviously, pairings are symmetric and therefore also $\pi(j)=i$ holds true. Unpaired bases are represented by  ``fixed points'' $\pi(i)=i $. With such conventions, the permutation is an {\it involution}, that is $\pi(\pi(i))=i$ for all $i=1,\ldots,L$.

\subsection{Irreducible diagrams}
\begin{figure}[h]
  \centering
   \includegraphics[width=3in]{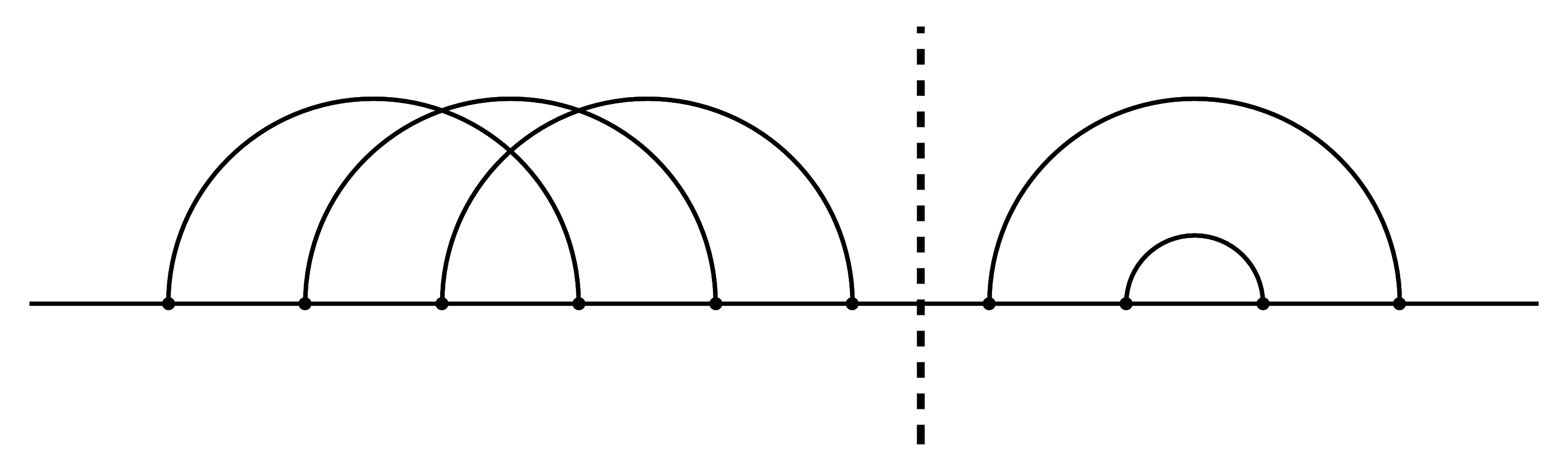}
  \caption{A reducible diagram can be decomposed in two disconnected components by cutting the backbone once (dotted line).}
  \label{f4}
\end{figure}

To decompose any RNA graph in its irreducible components, is sufficient to verify recursively whether it can be disconnected by cutting the backbone at any one point (see fig. \ref{f4}). 
In particular, we may use an electrostatic analogy where the pairings and the backbone are regarded as electrostatic field lines. We assign a positive charge $q=+1$ to every base where a pairing begins, (i.e.  with $\pi(i)>i$), a negative charge $q=-1$ to every base where a pairing ends, (i.e.  with $\pi(i)<i$),  and no charge, $q=0$, to every free base (i.e.  with $\pi(i)=i$):
\begin{equation}
q_i=sign\left( \pi(i)- i\right) \, , \quad i=1,\ldots,L  \, ,
\end{equation}
where $sign$ is the sign function (equal to 0 for vanishing argument). When the cumulative sum  $c_k=\sum_{i=1}^{k} q_i$ is zero, then all pairings that started before the $k$-th base also must have ended before the $k$-th base. In fact, the RNA segment up to the base $k$ (included) is charge neutral, and thus is loosely bound to the rest of the molecule (i.e. there are no unbalanced pairings to the left of $k$). By cutting the backbone just on the right of the $k$-th base, the molecule disconnects into two separate components. This procedure can be repeated all the way to the 3' end of the RNA molecule (up to $i=L$), and every time that the cumulative sum $c_k$ is zero, the graph can be disconnected by cutting the backbone  at the base $k$. The pseudocode implementing such procedure is in Algorithm \ref{A1}.
\begin{algorithm}
\caption{Decompose a diagram into irreducible components}
\label{A1}
\begin{algorithmic}[1]
\REQUIRE $\pi$ (a permutation involution)
	\STATE $L$=length($\pi$)
	\STATE	StartsAt=1;c=0
	\FOR{ i=1 to L}
		\STATE c=c+sign($\pi$(i)-i)
		\IF{ c is 0} 
			\STATE Print ``irreducible from '' StartsAt ``to'' i
			\STATE StartsAt=i+1;
			
		\ENDIF 	
    \ENDFOR
\end{algorithmic}
\end{algorithm}

\subsection{Nested diagrams}
The next essential tool is the identification of all nested components in the diagram. A diagram is said to contain a nested component if such a component can be removed by cutting the backbone at {\it two} points. The concept of ``nestedness'' is closely related to the concept of irreducibility. This can be illustrated by introducing the cyclic (right) shift-permutation $\sigma= (2,3,4,\ldots,L,1)$. Under such shift permutation, every site $i$ is mapped onto its right-neighbor $i+1$. Moreover,  the permutation is cyclic in the sense that the last base $i=L$ is mapped onto the first one $i=1$. By applying the shift permutation $\sigma$ a sufficient number of times, any nested component of the diagram can be translated to the right until its rightmost base touches $i=L$. Such a diagram is  reducible evidently. Therefore, one can identify all nested components of a diagram by simply identifying all the irreducible parts of $\sigma^k \cdot \pi$ for all $k=1,\ldots,L$. Such a procedure is implemented in Algorithm \ref{A2}.
\begin{algorithm}
\caption{To identify all irreducible and nested components}
\label{A2}
\begin{algorithmic}[1]
\REQUIRE $\pi$ (a permutation involution)
	\STATE $L$=length($\pi$)
	\STATE $\sigma=(2,3\ldots,L,1)$  (the cyclic shift permutation)
	\FOR{ i=1 to L}
		\STATE find (and output) all irreducible components of $\pi$ (use Algorithm 1)
		\STATE $\pi=\sigma \cdot \pi$;			
	\ENDFOR
\end{algorithmic}
\end{algorithm}

We note that all free bases are by definition also nested components, since it is possible to disconnect any free base $i$ by simply cutting the backbone at $i-1$ and $i+1$. Therefore, when considering diagrams that do not have any nested component, one can as well consider diagrams where all free bases are removed. 

The possibility of identifying all nested components in a RNA diagram, opens the way to a procedure that we defined as ``backbone renormalization'' in \cite{19}. It consists of replacing each nested component by a new type of backbone segment, called $z_g$, where $g$ is the genus of the nested component that has been replaced. To that objective, we briefly review \cite{19}  how to compute the genus of any diagram (nested or not, irreducible or not).

\subsection{The genus} 
The explicit evaluation of the formula $g=(p-l)/2$  can be performed efficiently by using the formalism of permutations. In this case, the number of pairings, which is simply half the number of paired bases, is given by
\begin{equation}
p=\frac{1}{2}\sum_{i=1}^{L} (1-\delta_{i \pi(i)} )
\end{equation}
where $\delta$ is the Kronecker delta function. The total number of loops can be obtained by counting the number of cycles $c$ of the permutation $\sigma \cdot \pi$ where $\sigma$ is the cyclic shift-permutation \cite{23}. One can easily verify that $c=l+1$, that is, among all cycles there is also a loop which contains the cyclic link from $i=L$ to $i=1$. We have:
\begin{equation}
g=\frac{p-c+1}{2} \, .
\end{equation}
The pseudocode to compute the genus of a permutation involution $\pi$ is in Algorithm \ref{A3}.
\begin{algorithm}
\caption{Compute the genus of the diagram $\pi$}
\label{A3}
\begin{algorithmic}[1]
\REQUIRE $\pi$ (a permutation involution)
	\STATE $L$=length($\pi$)
	\STATE $\sigma=(2,3\ldots,L,1)$  (the cyclic shift permutation)
	\STATE $\tau=\sigma \cdot \pi$
	\STATE $c$ =  number of cycles of $\tau$ 
	\STATE $p=( L -$ number of fixed points of($\pi$) )/2;			
	\STATE genus=$(p-c+1)/2$
\end{algorithmic}
\end{algorithm}

It is straightforward to verify also that the genus is an additive quantity both in the nested components and in the irreducible parts. More precisely, if the (reducible) diagram ${\cal D}={\cal D}_1 \cup {\cal D}_2$ is the sum of two irreducible components  ${\cal D}_1, {\cal D}_2$, then $g({\cal D})=g({\cal D}_1)+g({\cal D}_2)$. Analogously,  if the diagram ${\cal D}$ has a nested component ${\cal D}_1$, then again $g({\cal D})=g({\cal D}_1)+g({\cal D}_2)$, where ${\cal D}_2$ is the complement of ${\cal D}_1$ in ${\cal D}$. 

\subsection{Primitive diagrams}
\begin{figure}[h]
  \centering
  \includegraphics[width=2.8in]{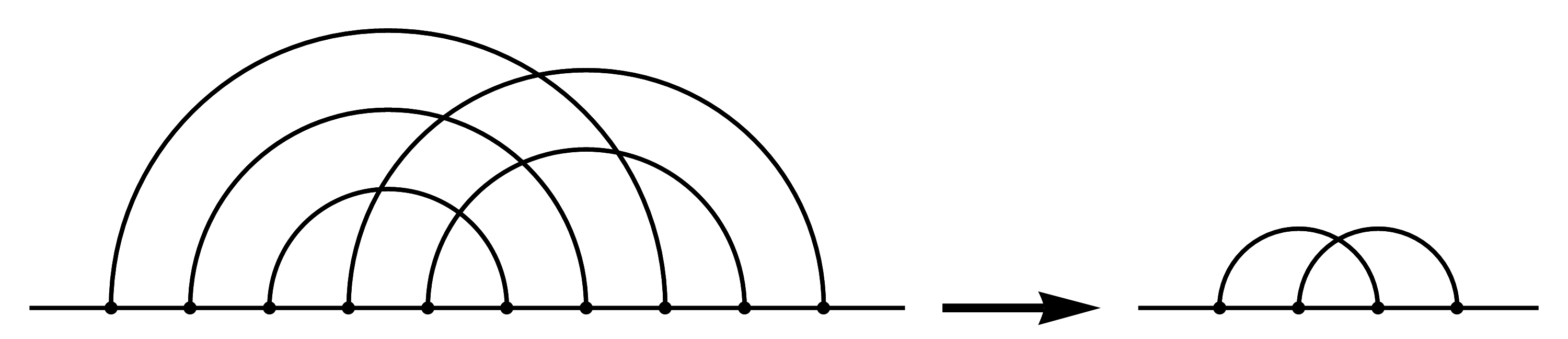}
  \caption{Arc renormalization: any sheaf of parallel arcs can be mapped into a single arc}
  \label{f5}
\end{figure}

The final requirement to characterize primitive diagrams is to collapse parallel pairing lines in the graph into a single one. We say that two lines (or arcs) are {\it equivalent} if they don't cross, and if they intersect exactly the same pairing lines (see Fig. \ref{f5}). 
A simple way to translate it into an algorithm is to define a primitive diagram as an irreducible, not nested diagram, with no stacked pairings. Any stacked pairing in $\pi$ corresponds to a cycle of length two for the composite permutation $\bar{\sigma} \cdot \pi$, where $\bar{\sigma}= \{2,3,4,\ldots,L,L\}$
is the {\it non-cyclic} (right) shift permutation. Therefore, a primitive diagram is represented by a permutation involution $\pi$ which is irreducible, not nested and such that  $\bar{\sigma} \cdot \pi$ does not contain any cycle of length two. A simple algorithm to ``renormalize'' nested arcs in a generic diagram, and to adsorb any nested planar diagram into renormalized backbones (of planar type only) is listed in Algorithm \ref{A4}.  
\begin{algorithm}
\caption{Primitive diagram}
\label{A4}
\begin{algorithmic}[1]
\REQUIRE $\pi$ (a permutation involution)
	\STATE flag=1
	\WHILE{flag=1}
		\STATE flag=0; $L$=length($\pi$)
		\STATE $\sigma=(2,3\ldots,L,L)$  (not-cyclic shift permutation)
		\FOR { i=1 to L}
			\IF[if there is a 2-cycle, then replace one of the two arcs with free bases.]{$\sigma ( \pi ( \sigma ( \pi(i))))=i$} 
				\STATE $\pi ( \pi (i))=\pi(i)$;  $\pi(i)=i$
			\ENDIF
		\ENDFOR
			\STATE counter=1	
			\COMMENT {Relabel the sequence, while skipping all fixed points}				
			\FOR{ j=1 to L}
				\IF {$\pi(j)=j$}
				\STATE label(j)=0
				\ELSE
				\STATE label(j)=counter; counter++
				\ENDIF
			\ENDFOR
			\FOR{ j=1 to L}
				\IF{$\pi(j)$ is not equal to $j$}					
				\STATE $\pi_{new}$(label(j))=label($\pi$(j))
				\STATE flag=1
				\ENDIF
			\ENDFOR	
		\STATE $\pi=\pi_{new}$
	\ENDWHILE
\end{algorithmic}
\end{algorithm}

\subsection{The renormalized crossing number}
We conclude this section by providing an algorithm to compute the crossing number and the renormalized crossing number of a generic diagram. The crossing number is the lowest number of crossing points among arcs in the diagram. Like the genus, the crossing number is also an additive quantity with respect to nestedness and reducibility. In simple words, the crossing number of any reducible (or nested) diagram ${\cal D}={\cal D}_1 \cup {\cal D}_2$  is  $N_c({\cal D})=N_c({\cal D}_1)+N_c({\cal D}_2)$.  However, as we have discussed previously the crossing number is not invariant under arc-renormalization: for instance, the crossing number of the diagrams in Fig. \ref{f5} is $N_C=6$ for the graph on the left and $N_C=1$  for graph on the right. Algorithm \ref{A5} parses the RNA permutation involution $\pi$ and for each arc ($\pi(i)>i$), first it counts the number of intersecting arcs between $i$ and $\pi(i)$, and then removes it. While Algorithm  \ref{A5} works for any graph, including primitive ones, in our thermodynamic model eq.~(\ref{eq2}) only the renormalized crossing number of a graph is necessary. As explained in the introduction, the rationale is that  RNA databases do not show a preference for short vs. long helices for same-genus pseudoknots, in addition to the enthalpic contribution. Algorithm \ref{A6} outlines the pseudocode for computing the renormalized crossing number.

\begin{algorithm}
\caption{Compute the crossing number of $\pi$}
\label{A5}
\begin{algorithmic}[1]
\REQUIRE $\pi$ (a permutation involution)
	\STATE $L$=length($\pi$)
	\STATE CrossingNumber=0
	\FOR{i = 1 to L}
		\IF[rising arc] {$\pi(i)>i$} 
			\STATE a=i
			\STATE b=$\pi(i)$
			\FOR[for each base inside the arc]{j = $a+1$ to $b-1$}
				\IF{$\pi(j)$>b}
					\STATE CrossingNumber++
				\ENDIF
			\ENDFOR
			\STATE  $\pi(\pi(i)) = i$
			\STATE $\pi(i)=i$		
		\ENDIF
	\ENDFOR
\end{algorithmic}
\end{algorithm}

\begin{algorithm}
\caption{Compute the renormalized crossing number $N_c$ of $\pi$}
\label{A6}
\begin{algorithmic}[1]
\REQUIRE $\pi$ (a permutation involution)
	\STATE Use Algorithm A2 to find all $m$ irreducible and nested components ${\cal D}_i$ of  $\pi$.
	\STATE $N_c$=0
	\FOR{i = 1 to m}
		\STATE Use Algorithm A4 to compute the primitive diagram ${\cal D}'_i$ of ${\cal D}_i$.
	    \STATE Use Algorithm A5 to compute the crossing number $n_c$ of ${\cal D}'_i$.
		\STATE $N_c=N_c+c$
	\ENDFOR
\end{algorithmic}
\end{algorithm}
By using these algorithms to compute the genus and the crossing number of a graph, it is possible to perform a Monte Carlo sampling of graphs of the system analogous to  ref. \cite{18}, using the energy of eq. (\ref{eq2}). The full implementation of the algorithm and the fitting of the genus and crossing number penalties will require additional work which will be presented in a forthcoming paper.

\section{Conclusions}
In addition to a topological chemical potential coupled to the genus, we propose to add a term proportional to the renormalized crossing number of a RNA graph to the energy function of pseudoknotted RNAs. Such a procedure requires the systematic evaluation of the primitive diagram of any RNA secondary structure, with or without pseudoknots. 
In turn, that can be expressed naturally with the  formalism used in matrix quantum field theory to renormalize Feynman diagrams. We discussed two levels of renormalization: the backbone and the arc renormalization, leaving the vertex renormalization to a future paper. The latter is helpful not only to collapse the RNA diagram into simpler ones, but can be used for building new diagrams with higher topological complexity from simpler ones. We are currently implementing the Monte Carlo algorithm with the modified energy function to predict RNA structures. In order to do so, it is necessary compute the change in the genus and in the renormalized crossing number of a graph upon addition or removal of a helical fragment (equivalent to a single pairing in the primitive graph). The incremental change of the genus was described in ref. \cite{17}, and the change of crossing number will be discussed in a forthcoming paper. However, as already pointed out in  ref. \cite{18}, all these algorithms based on topology do not take into account the geometry of the molecule, and in particular, many of its predictions are plagued by steric clashes. The next challenge will be to include the steric constraints at each step of the Monte Carlo procedure.  
 
\vspace{1cm}
\noindent
\underline{Acknowledgments}:
One of us (H.O.) would like to thank Joel Hass for illuminating discussions, and the Physics Department of UCSB for its generous hospitality during part of this work.

\end{document}